\newcommand{\simle}{\mbox{$\stackrel{<}{_{\sim}}$}}

\documentstyle[12pt,emulateapj,epsfig]{article}
\lefthead{Tuthill et al.}
\righthead{IRC\,+10216 proper motions}

\begin{document}
%_____________________________________TITLE PAGE___________________________
%\title{Milliarcsecond motions of IRC\,+10216's inner dust nebula: smoke signals
%       from a dying carbon star.} 
\title{Smoke Signals From IRC\,+10216\\
1. Milliarcsecond Proper Motions of the Dust} 

\author{P. G. Tuthill\altaffilmark{1,3}, J. D. Monnier\altaffilmark{2,3}, 
W. C. Danchi\altaffilmark{3} and B. Lopez\altaffilmark{4}
} 

\altaffiltext{1}{Chatterton Astronomy Department, School of Physics, University 
of Sydney, NSW 2006, Australia }
\altaffiltext{2}{Smithsonian Astrophysical Observatory, MS42, 60 Garden Street, 
Cambridge, MA, 02138 , USA}
\altaffiltext{3}{Space Sciences Laboratory, University of California, Berkeley,
Berkeley,  CA  94720-7450 }
\altaffiltext{4}{Observatoire de la C\^ote d'Azur, Departement Fresnel UMR 6528,
BP 4229, F-06034 Nice Cedex 4, France}

%\email{gekko@physics.usyd.edu.au, 
%       jmonnier@cfa.harvard.edu, 
%       wcd@ssl.berkeley.edu, 
%       lopez@obs-nice.fr}
%

%____________________________________ABSTRACT PAGE_________________________
\begin{abstract}

The results of a 7-epoch interferometric imaging study, at wavelengths
in the near-infrared K-band, of the carbon star IRC\,+10216 are presented. 
The use of non- and partially-redundant aperture masking techniques 
on the 10-m Keck-I telescope has allowed us to produce images of the innermost 
regions of the circumstellar dust envelope with unprecedented detail.
With roughly twice the resolving power of previous work (\cite{Weigelt97}; 
\cite{Weigelt98}; \cite{HB98}), the complex asymmetric structures reported 
within the central \mbox{0 \farcs 5 ($\sim$20\,R$_\star$)} have been imaged 
at the size scale of the stellar disk itself ($\sim$50\,mas).
A prominent dark lane at a position angle of approximately $120^\circ$
is suggested to be an optically thick disk or torus of dust which could help to
explain IRC\,+10216's well-known bipolarity at a position angle of $\sim20^\circ$.
Observations spanning more than a pulsational cycle ($\sim$638\,days) have 
revealed significant temporal evolution of the nebula, including the outward 
motion of bright knots and clumps.
Registering these displacements against the compact bright core, which we
tentatively identify as marking the location of the star, has allowed us to 
determine the apparent angular velocity at a number of points.
The magnitudes of the proper motions were found to be in agreement with 
current estimates of the stellar distance and radial velocity.
Higher outflow speeds were found for features with greater separation from 
the core.
This is consistent with acceleration taking place over the region sampled by the 
measurements, however alternate interpretations are also presented. 
Although a number of changes of morphology were found, none were clearly 
interpreted as the condensation of new dust over the pulsation cycle. 
Unfortunately, ambiguities associated with the true three-dimensional nature
of the nebula weaken a number of our quantitative and qualitative conclusions.

\end{abstract}


%_______________________________________INTRODUCTION_______________________
\section{Introduction}

The extreme carbon star IRC\,+10216 is a classic example of a red giant caught 
in the act of evolving into a planetary nebula.
Its relative proximity, high infrared luminosity, and abundance of molecules
found in its dense outflow has resulted in a barrage of observations by astronomers, 
working across the spectrum, but particularly in the infrared and 
millimeter/sub-millimeter.
Despite all this attention, a good model of what is happening in the innermost 
regions where the stellar outflow is born and accelerated is still sorely lacking.

Numerous studies of molecular lines in the outer envelope (e.g. \cite{BT93}) have 
revealed a spherically expanding outflow, a finding which was beautifully confirmed
with deep $B$ and $V$ band images of the dust shell in ambient scattered galactic 
light (\cite{Mauron99}).
However this spherical symmetry, a characteristic of most red giant winds, will 
likely be broken as the IRC\,+10216 evolves into a planetary nebula, most of which
are elongated or bipolar (e.g. \cite{ZA86}).
The pronounced asymmetry in the innermost regions of the envelope of IRC\,+10216
reported by numerous high-resolution imaging experiments (most recently
\cite{Weigelt97}; \cite{coolstars98}; \cite{Weigelt98}; \cite{HB98}) and also
polarization studies (\cite{TDW94}; \cite{KW94}) suggests that the onset of this
aspherical flow has already begun, probably within the last few hundred years.
With a privileged vantage onto such a brief yet important period in the evolution
of a low to intermediate mass (initial mass $\sim 3$\,--\,5\,M$_\odot$ \cite{Guelin95}) 
star, high resolution observations are crucial in distinguishing between the many 
competing models for the physical mechanisms underlying the onset of asymmetry 
in the birth of a planetary nebula.

In this paper, we present a 7-epoch diffraction-limited imaging study of the inner 
dust shell of IRC\,+10216 in the near-infrared K-band.
Although some interpretation of the morphology of the images is given, 
full radiative transfer modelling results are beyond the scope of this report, 
and will be presented in a second paper.
Instead, we emphasize here the detection and measurement of the motion of 
features presumably embedded in the outflow.
Although proper motions have been reported for near-infrared images of dusty Wolf-Rayet 
shells (\cite{wr104}; \cite{wr98a}), the two order-of-magnitude slower winds around 
Asymptotic Giant Branch (AGB) stars result in the requirement of longer time bases, 
and extremely high fidelity mapping schemes.

%________________________________OBSERVATIONS/RESULTS_________________________
\section{Observations and Results}

\subsection{Observations}

Diffraction-limited images of IRC\,+10216 were obtained at seven separate epochs 
with the Keck~I telescope, with dates and other observing details given in Table~1.
Observations used the technique of aperture masking interferometry, by which starlight 
from the primary mirror is selectively blocked, with only a few regions of the pupil
allowed to contribute to the final image.
For observations of bright compact objects, of which IRC\,+10216 is a prime example,
the methods of sparse-pupil interferometry have been shown to be fully competitive
with, or superior to, other techniques such as speckle interferometry.
%The primary reason for this somewhat counter-intuitive result of gaining in 
%signal-to-noise by throwing away light stems from the corresponding lowering of the 
%redundancy (the number of times a given vector baseline is repeated within the pupil).
%Apart from tailoring the pupil geometry, aperture masking has much in
%common with speckle interferometry in that sequences of rapid-exposure data frames
%are analysed to extract information on the complex visibility of the source --
%which is related to the image by a Fourier transform. 
Statistical methods based on the maximum-entropy technique (\cite{devinder}) have 
been used to recover maps from the complex visibility data; however alternate methods 
such as the CLEAN algorithm (\cite{clean}) produced similar results.
Data reduction and analysis procedures were also tested by observing a number of
test objects, such as known binary stars, on each night. 
A detailed description of the Keck aperture masking experiment covering the
observational techniques, data reduction, and image reconstruction can be
found in Tuthill et al.~(2000).

%%%%%%%%%%%%%%%%%%%%%%%%%%%%
% Table - Observing Log    %
%%%%%%%%%%%%%%%%%%%%%%%%%%%%
\begin{deluxetable}{cllll}
\tablewidth{0pt}
\tablecaption{Journal of Interferometric Observations.}
\tablehead{
\colhead{Epoch} & \colhead{Date} & \colhead{Mask} & 
\colhead{Filter\tablenotemark{1}} & \colhead{Phase\tablenotemark{2}}}
\startdata
1 & 1997 Jan 29 & Golay15 & kcont	& 0.72 \nl
1 & 1997 Jan 29 & Annulus & kcont	& 0.72 \nl
2 & 1997 Dec 16 & Annulus & kcont	& 1.23 \nl
2 & 1997 Dec 16 & Golay21 & kcont,ch4 	& 1.23 \nl
2 & 1997 Dec 18 & Golay21 & kcont	& 1.23 \nl
3 & 1998 Apr 14 & Golay21 & ch4		& 1.41 \nl
3 & 1998 Apr 15 & Annulus & ch4		& 1.41 \nl
4 & 1998 Jun 04 & Golay21 & ch4		& 1.49 \nl
5 & 1999 Jan 05 & Annulus & ch4		& 1.83 \nl
5 & 1999 Jan 05 & Golay21 & ch4		& 1.83 \nl
6 & 1999 Feb 04 & Golay21 & ch4		& 1.88 \nl
6 & 1999 Feb 04 & Annulus & ch4		& 1.88 \nl
7 & 1999 Apr 25 & Annulus & k,ch4	& 2.00 \nl
7 & 1999 Apr 25 & Annulus & ch4		& 2.00 \nl
\enddata
\tablenotetext{1}
{
Filter Wavelength Information ($\mu$m):\nl
\begin{tabular}{ccc}
       & $\lambda_0$ & $\delta \lambda$ \nl
kcont  & 2.260 & 0.053 \nl
ch4    & 2.269 & 0.155 \nl
k      & 2.214 & 0.427 \nl
\end{tabular}
}
\tablenotetext{2}{Stellar Phases from \cite{MGD98}}
\end{deluxetable}

At each epoch in Table~1, a number of observations of IRC\,+10216 were made, often
with quite distinct experimental setups.
Three different aperture masks were used over the course of the project, with the 
non-redundant Golay-type masks passing only a few percent, and the partially-redundant 
annulus mask passing around ten percent of the unobstructed pupil.
Three different filters were also employed at various epochs, with the bandpass 
characteristics given in Table~1.
This diversity of observing parameters, while in part representing experimental 
evolution, allowed us to tailor the observations to conditions and specific 
requirements on a given night.
However, no systematic differences were found when comparing maps from annulus and
Golay data, and more significantly, from any of the different filters used
(filters differ in bandwidth, but have similar center frequencies; c.f. Table~1).
Thus for the remainder of this work, all maps taken at one observing epoch are
treated as measuring the same quantities.

Two limitations of the interferometrically reconstructed maps should be
mentioned.
Firstly, the absolute photometry, or surface brightness scale in the maps, is
difficult to calibrate with any great accuracy.
For this reason, images shown have fluxes scaled relative to the peak intensity in
each map, and discussion will refer to these relative fluxes.
Secondly, the closure phase method does not deliver absolute positional 
information, and the center of each map has been chosen to be the location
of the brightest pixel.

\subsection{Morphology of IRC\,+10216}

\begin{figure*}[p]
\begin{center}
\epsfig{file=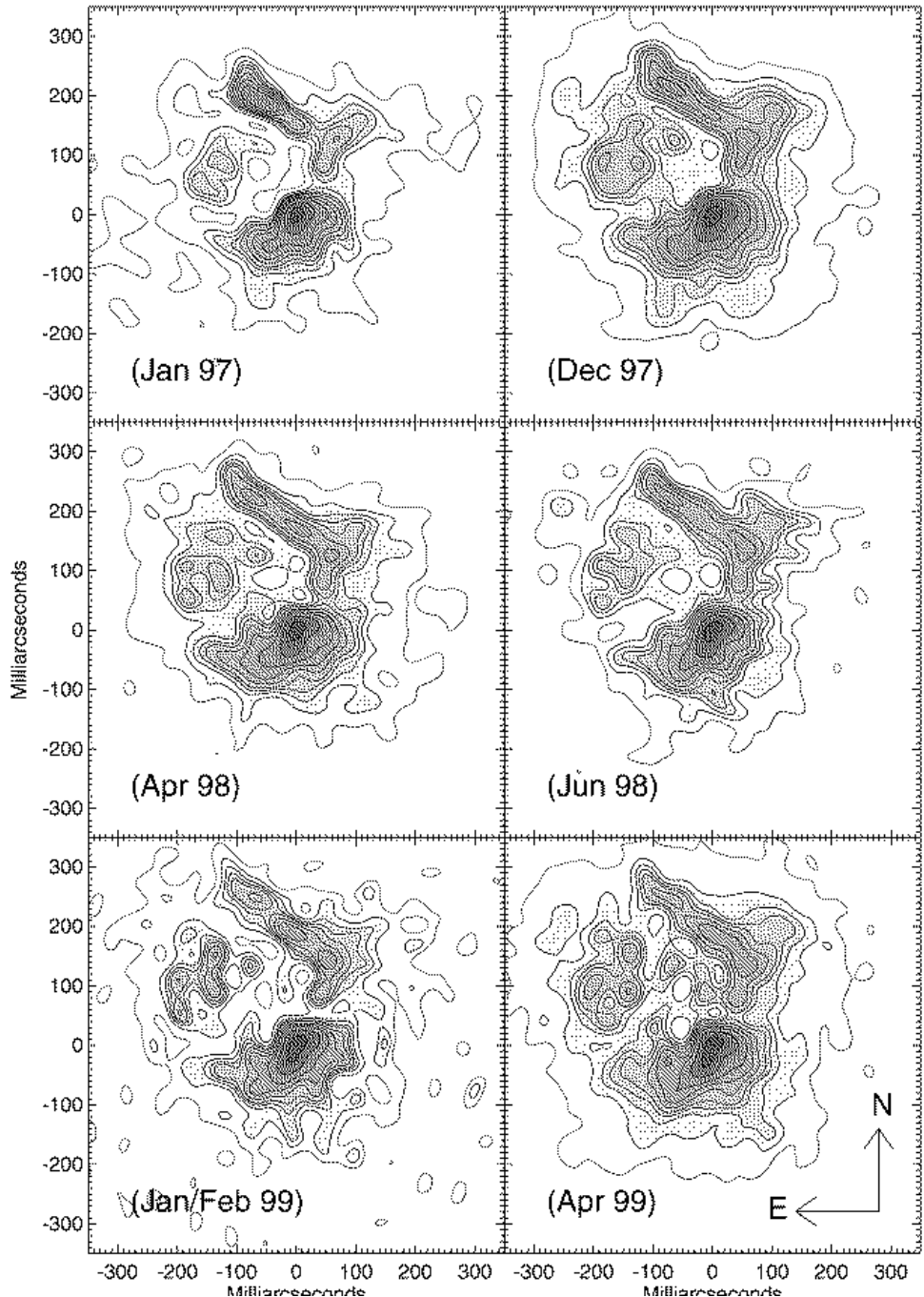,height=20cm}
\end{center}
\caption{ \label{fig:6maps}
Image reconstructions of IRC\,+10216 in the near-infrared K~band taken 
over a period from 1997 January to 1999 April.
The contour levels, at  1,\,2,\,3,\,4,\,5,\,7,\,10,\,15,\,30 \& 70\,\% of 
the peak pixel, highlight the extended spatial structure.
Each map presented is the average of between one and six individual maps, 
each taken at different times, and often with differing observing 
geometries, filters and seeing conditions. 
By performing a noise-weighted average over a number of maps, random and
systematic noise in the images could be suppressed for observations at
a single epoch. 
The exception is the Jan/Feb~99 map, where data were averaged over the 
two separate (but closely spaced) epochs.
This was found to be necessary as poor data quality resulted in noisy 
maps over this period.}
\end{figure*}

Figure~\ref{fig:6maps} shows maps of IRC\,+10216 from data spanning a period from
1997 January to 1999 April.
In this figure, maps shown are the noise-weighted average over all image 
reconstructions from a given epoch (or, in the case of Jan/Feb~99, pair of epochs).
Simple inspection shows that the basic morphology of the inner nebula surrounding
IRC\,+10216 has been very well established in the K~band.
From the most prominent structures down to relatively minor features at only a few 
percent of the peak surface brightness, a high degree of consistency is found 
in the sequence of images.
Furthermore, the early epoch images shown in Figure~\ref{fig:6maps} are also in 
excellent agreement with  recently published near-infrared images of 
Weigelt et al.~(1998) and Haniff \& Buscher (1998) (We refer collectively to these 
two papers as ``WHB'' hereafter).
%Lowering the resolution of the Keck images by convolving with an Airy disk 
%appropriately scaled for the smaller telescopes used (6\,m and 4\,m respectively), 
%produced maps almost identical to those reported earlier.

In order to proceed with a quantitative analysis of the maps, a simple descriptive
model has been used to identify and label features. 
WHB have labelled compact features A through D, with Weigelt et al.~(1998) adding 
E and F which seem much less distinct.
With the considerably higher resolution available from the Keck, these compact knots
have in most cases been resolved into more complicated structures, and therefore a 
different approach is taken in describing the images, which we relate to the earlier 
schemes.

\begin{figure*}[htbp]
\vspace{0cm}
\begin{center}
\epsfig{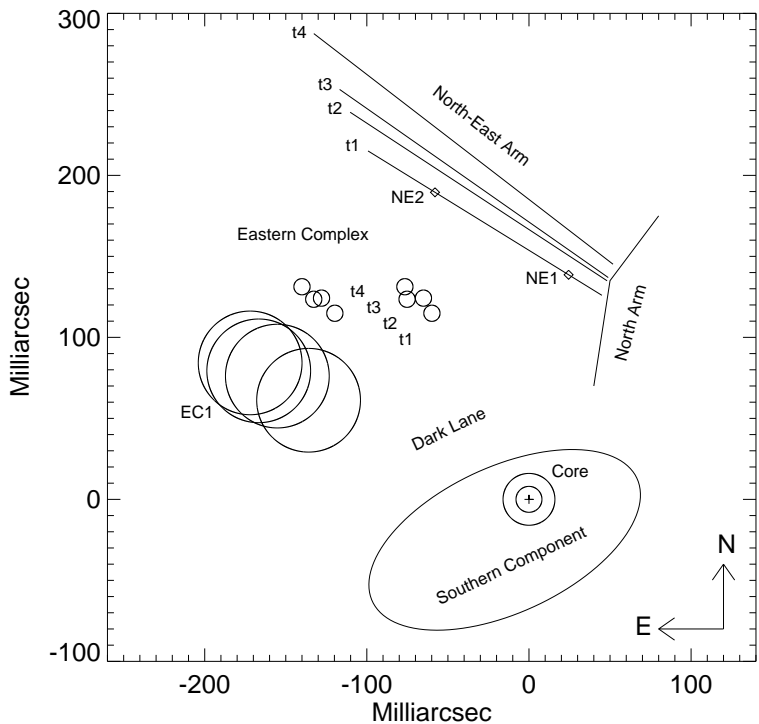}
\end{center}
\caption{ \label{fig:cartoon}
Skeleton diagram of features identified in high resolution images of
IRC\,+10216, drawn to scale and registered so that objects appear relative
to the location of the map peak at (0,0), denoted by the symbol + above. 
A number of other features, such as the Southern Component, the
North Arm, the North-East Arm and the Eastern Complex labelled
above are described further in the text.
Taken relative to the peak, these features were found to move with time, 
and to avoid unnecessary clutter in the diagram, we have plotted the best-fit
locations at only four (rather than seven) epochs, labelled t1 through t4 above. 
Epochs t1 and t2 correspond to 1 and 2 (i.e. Jan 97 \& 
Dec 97) from Table~1, while t3 and t4 are the result of averaging over 
epochs $3 + 4$ and $5 + 6 + 7$ respectively.
Note that this was done for the purposes of simplifying this illustration only, 
and data averaged were never more than three months apart (as for epoch t4).
The computer-fit locations of the North-East Arm and elements in the
Eastern Complex are clearly migrating away from the origin
through epochs t1 to t4.
Points labelled NE1 and NE2 identify fixed locations towards the ends
of the North-East Arm, and are used further in the quantitative analysis
described in the text. }
\end{figure*}

A skeleton diagram of our model, based on analysis of images from Figure~\ref{fig:6maps},
is given in Figure~\ref{fig:cartoon}.
The location of the brightest feature in all maps, the relatively compact 
core, appears as a + surrounded by rings showing its approximate extent
(Feature A of WHB).
This core appears somewhat offset to the North-West from the center of an elongated,
roughly elliptical region (Features E \& F of Weigelt et al.~(1998)).
We refer to the Core and these immediate surroundings as the Southern Component in 
Figure~\ref{fig:cartoon}.
The next most prominent structure in the maps is the linear extension
to the North and North-East, which we have split into two; the shorter
North Arm (WHB Feature C) and the more prominent, elongated, 
North-East Arm (WHB Feature B).
The important North-East Arm we have modelled as a linear ridge of emission.
Displaced to the East between these other features is a dimmer structure,
consisting of multiple peaks we have labelled the Eastern Complex 
(WHB Feature D).
Although signal-to-noise was not uniformly high enough to be assured of high-fidelity
reconstructions of the Eastern Complex, for the purposes of characterizing the structure, 
we have labelled the most southerly portion Cloud EC1 (usually a little brighter and
appearing like a ring or three peaks in a triangle), while in the more northern 
section we have been content to simply tag the location of the two most prominent peaks.

\subsection{Proper Motions and Modelling}

Having established in the previous section a framework for describing the appearance 
of the maps of IRC\,+10216, we now proceed to examine the data for changes 
over the seven epochs.
The easiest thing to look for is a change in the relative positions of the components.
This was done with the use of a computer program which found the best-fit location of
model components describing four features: the North-East Arm, EC1, and two minor
peaks in the Eastern Complex. 
As our maps have no associated astrometry, the registration between separate images 
is unknown and we have used the bright compact core as the fixed point
against which to measure any motions.
Note that fits were not made directly to the averaged images presented in 
Figure~\ref{fig:6maps}, but to the full 7-epoch dataset with multiple separate 
images at each epoch (comprising a total of 25 maps).
This allowed errors to be determined on the locations of the features from the 
apparent spread of values over different maps.

All four features tracked were found to exhibit widening separations from the core
as a function of time.
This can be verified by visual inspection of the images of Figure~\ref{fig:6maps}
where the Northward motion of the North-East Arm is readily apparent without any
assistance from computer model-fits.
Motions of the North-East Arm and elements in the Eastern Complex including EC1
are shown in Figure~\ref{fig:cartoon}.
To avoid the confusion of seven sets of features on one plot, the motion is shown 
averaged into four time-intervals (labelled t1 -- t4), as described in the caption.

In addition to the motions of certain features, other changes are apparent with time
in Figure~\ref{fig:6maps}, at a level of significance well above the level of noise
in the maps.
Two of these in particular are worth highlighting.
The North-East Arm which in 1997 January ends in a bright, fairly compact knot, evolves 
through a stage where it is of fairly uniform brightness along its length (1998), and
in the final measurements (1999) the extreme end of the arm has dimmed considerably.
The second interesting change concerns the central core, which in 1997 January appears 
fairly circular and compact, but by 1999 April clearly exhibits an extension to the 
South-East.
Further discussion of these points is given in Section~3 below.

\subsection{Outflow Velocities} 

Having established the presence of motion between the components, it is possible
to characterize the apparent outflow of material by watching the features 
which are presumably embedded within it.
We restrict our attention to the North-East Arm and EC1, as the more minor 
features are too near to the map noise level to make useful tag points for the flow.
The first important question to be addressed is the assumption as to
the origin of the flow.
WHB present arguments that their component A, the compact core, be identified
as the star itself. 
Certainly, features identified to be moving do appear to be moving away from 
a point to the South, and in our subsequent analysis we have assumed radial
divergence from the compact core.
However, there is no guarantee that any of the map features must be the stellar 
disk, which may lie behind some optically thick region, as is discussed further
below.
 
It was fairly straightforward to project the motion of EC1 onto a vector 
beginning at the origin (the brightest pixel in the core; see 
Figure~\ref{fig:cartoon}).
Displacement along this vector with time then gives a velocity.
Motion of the North-East Arm, however, is not so easy to quantify.
The minimum possible apparent velocity consistent with the time-sequence of images
would have material in the arm moving to the North-West, perpendicular
to its length.
While giving a lower bound on the possible velocity, this motion is not consistent 
with spherical outflow from our defined origin, and furthermore does not really 
describe the data as the arm would also need to grow in length. 
Instead, we have labelled points NE1 \& NE2 near the beginning and end of the arm 
respectively, and we measure the {\em radial} motion of these points from the 
origin.
Since the Arm has been fit as a linear ridge, two points are sufficient to 
completely determine its motion.
Thus our three trace points in the flow are EC1 and the two points NE1 \& NE2 on 
the North-East Arm, all of which are assumed to be radially diverging from the Core. 

\begin{figure*}[htbp]
\begin{center}
\epsfig{file=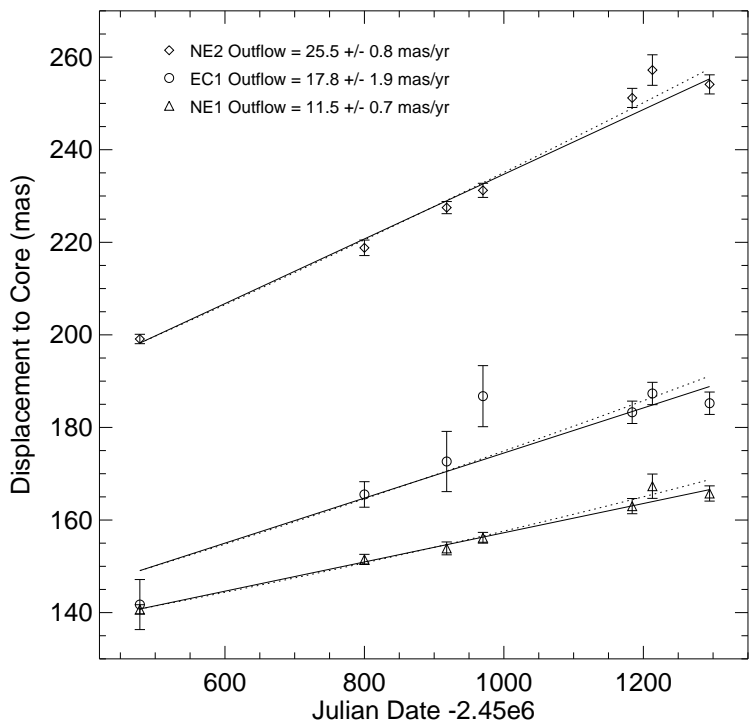,height=10cm,angle=90}
\end{center}
\caption{ \label{fig:outflow}
Plot of the displacement from the origin (brightest map pixel) as a function
of observing Julian date for model features identified in the maps of IRC\,+10216. 
A radially divergent flow from the origin has been assumed.
Tracking three features (points NE1, NE2 and EC1 -- see Figure~2 and
text) over seven epochs (see Table~1) reveals linear relationships of displacement
versus time, as is especially apparent for the well-defined high signal-to-noise
North-East Arm points NE1 and NE2.
The velocity of the flow at each point has been obtained from the slope 
of the linear fits (solid lines) and is given in the figure Key.
Also over-plotted are dotted lines showing an acceleration law of 3.4\,mas.yr$^{-2}$.
This is for illustrative purposes only, and does not represent a fit to 
curvature of the displacement versus time data given here. 
See text for further details.}
\end{figure*}

Plotted in Figure~\ref{fig:outflow} are the displacements from the origin of each 
trace point over the seven epochs.
For the well-determined points NE1 \& NE2 lying on the high signal-to-noise
North-East Arm, the points can be seen to describe a very well defined linear 
relationship with time, which is echoed, with more scatter, by the EC1 data.
Velocities may be computed from the slope of a least-squares fit to the data, with 
resulting apparent outflow rates given in Figure~\ref{fig:outflow}.

\subsection{Outflow Velocity Dispersion}

The most direct way to observe an accelerating body is to follow its change of
speed with time.
Unfortunately, the error bars are too great, and the time-span of observation too
short to make this strategy worthwhile for data in Figure~\ref{fig:outflow}.
A uniform radial acceleration  of 3.4\,mas.yr$^{-2}$ (see below) has been 
over-plotted (dotted line) illustrating the difficulty of measuring such a small 
curvature of the velocity law.
However, some information on the flow dynamics can be inferred from examination
of the velocity field derived from the maps.
Specifically, the greater the distance from the origin, the higher the outflow speed.
Three possible explanations for this finding are described below.

The acceleration of material over the region sampled by the measurements may
have been detected.
This acceleration can be easily visualized from the motion of the North-East Arm: 
the ridge-lines of best fit are not parallel causing the Arm to tilt Northward 
with time.
In order to do this, the far end of the arm (NE2) must be moving faster than
the near end (NE1).
The three flow-velocity data points from Figure~\ref{fig:outflow} are consistent
with a uniform radial acceleration of $3.4\pm0.5$\,mas.yr$^{-2}$ starting at a
radius of $80\pm20$\,mas.
As the principle acceleration mechanism is expected to be radiation pressure 
from the central star, a uniform acceleration law would not be expected.
However, in a region as clearly anisotropic as the inner dust shell of IRC\,+10216, 
flows may be complex and more realistic models will require greater efforts 
both in modelling and in recovery of longer time-baseline data.

A second possible explanation for the observed velocity field arises from the
ambiguities associated with the projection of three-dimensional structure
onto the plane of the sky.
Such projected motions and velocities may give a misleading view of the true
flow structure.
To take an extreme example, the North-East Arm may, in reality, be a fragment of a 
circular arc, all of which is at a uniform distance from the star and moving
at a uniform velocity.
However, when viewed from a relatively acute projected angle, apparent differences
in separation and velocity will be recorded.
For a roughly spherical distribution of clumps at a uniform flow speed, projection 
effects alone would result in slower apparent speeds being observed closer to
the star, mimicking an acceleration.
The uniform `acceleration' law given above is appropriate for such a scenario, where
the velocity will be proportional to the displacement from the origin.

Finally, a third possibility which presents itself is that faster material is
found further out (e.g velocity law $V \propto R$), but without acceleration
taking place over the sampled region.
The most obvious origin for such a flow would be a stellar eruption ejecting 
material with a range of velocities at some point in the past, allowing faster 
clumps to move further from the star.
Under this scenario, back-projection of the flow through time, based on the NE1 
\& NE2 velocities, implies that the North-East Arm was created in an event around 
JD~2449000 (about 1993 January) originating from a point some 90\,mas from
the core. 
Ideally, the assumption of a single expulsion event could be tested with three
or more clumps by seeing if they appear to be diverging from a single point 
in space and time.
Unfortunately, the errors on the EC1 flow are too large to offer a meaningful
constraint, although a common origin for NE1, NE2 \& EC1 does fall within the 
errors.

In summary, although acceleration stands as a strong candidate, the ambiguities
of interpretation make it impossible to claim a clear detection.
Indeed, some combination of all three effects discussed above may be needed
to account for the true angular velocity field.
This highlights the need for high quality imaging over more extended periods to
follow clumps from birth to dispersal in the extended shell.
Despite the uncertainties, a plane-of-the-sky assumption represents a simplest
case and should give, at the least, valid lower bounds to the velocity 
determinations discussed here.

%___________________________________DISCUSSION________________________
\section{Discussion}

Taking the Southern Component and the North-East Arm as the dominant bright
structures in our images, the bipolarity at position angle $\sim20^\circ$ 
reported throughout the literature is confirmed here.
However, at very high resolution, complex and clumpy structures are revealed 
which do not conform immediately to any simple axially symmetric models.
Perhaps one of the most striking features of the maps of Figure~\ref{fig:6maps}
is the Dark Lane (see Figure~\ref{fig:cartoon}), around which all the major 
structures are distributed.
At all epochs, the flux level in this hole, in close proximity to all
the brightest knots of emission, is consistent with a surface brightness of
$\simle 1 - 2$\% of the peak: similar to the noise level in the maps.
That such a dark region could exist in the heart of the nebula argues for
the presence of considerable material in the line-of-sight, leading to high
levels of obscuration.
If this is the case, the interpretation of the maps becomes more difficult.
A bright knot might be either a hot clump of dust, or a ``window'' in the
dust allowing a glimpse of the hot inner regions. 

The modelling of the dust shell is a current work in progress, with the most
promising interpretation being that the dark central band is a dusty torus
or equatorial density enhancement, which we can see tilted towards us to the
South revealing the inner hot regions (Southern Component) and possibly the
star (Core).
Clumpy features embedded in the outflow, such as the North-East Arm and
Eastern Complex, might have their origins in enhanced dust formation occurring
above slowly-evolving massive convective features (\cite{Weigelt98}) or magnetic 
spots (\cite{Soker99}) on the stellar surface.
Interestingly, no changes observed over the seven epochs, which cover more than
one pulsational cycle ($\sim$638\,days), gave clear evidence for new dust
nucleation (discounting the elongation of the Core, discussed below).
This is in accord with model dust shells (e.g. \cite{Winters95}) which show 
new layers of dust forming on timescales longer than one pulsation period.

Until further modelling can be completed, we confine further discussion 
here to the motion of material in the outflow.
For features such as the North-East Arm, it seems most likely that motions
detected are simply the displacement of emitting material, and not some more
exotic scenario such as the motion of a viewing hole in the dust, or a warm 
spot caused by a `searchlight beam' where the star shines through a moving 
window in the dust.
The North-East Arm does exhibit some common-sense characteristics reinforcing this
view, such as the flux from the bright knot at its extreme end, initially bright, 
fading as it moves outwards and presumably further from the central star 
(see Section~2.3).

Perhaps the greatest uncertainties affecting the interpretation are the
questions as to the origin and direction of the flow, and the three-dimensional
structure of the nebula.
Although Figure~\ref{fig:cartoon} does appear to show that features are moving
away from a point in the general location of the Southern Component, it was not
possible to pin this down precisely.
Worse, it was not possible to tell if the Southern Component and Core were moving
also: it may be the case that all features are diverging from the star which 
lies obscured behind the dark central band.
A number of previous authors (WHB, \cite{KW94}) have claimed that the central
star is visible in near-infrared.
The compact feature we have labelled the Core has an approximate angular size of 
$\sim50$\,mas -- close to the expected angular diameter of the stellar photosphere 
(\cite{bigpaper94}; \cite{JDM_thesis}).
However the progressive elongation of the Core through time (see Section~2.3; 
Figure~\ref{fig:6maps}) is not easy to reconcile with this view.
It is possible that the South-East extension we see is simply the newest 
condensation of dust moving from the star, however for this to be the case
the dust must be forming extremely close ($\simle 2R_\odot$) to the photosphere.
The inner radius of the dust shell has been recently estimated at $4.5R_\odot$
(\cite{Groen97}) and $6.8R_\odot$ (\cite{JDM_thesis}).
It may also be possible that a binary companion is playing some role in this
inner distortion, although it is highly unlikely that direct light from a 
main-sequence star could have been seen.
In this work, the assumption has been made that the Core does mark a fixed 
location associated with the star, with extra structure in the Southern 
Component arising from emission and/or partial obscuration from the inner 
boundary of the dust shell.
Alternate scenarios, in which the Core may also be moving, lead to modification 
of the derived velocities by up to a factor of 2.

The gas outflow velocity at large distances from the star is well established
from CO line profile studies to be 14.5\,km.sec$^{-1}$, however as calculated by
Groenewegen (1997), the dust will drift through the gas at 3\,km.sec$^{-1}$ 
resulting in a dust outflow speed of 17.5\,km.sec$^{-1}$.
Combining this with our maximum angular velocity of 25.5\,mas.yr$^{-1}$ from 
Figure~\ref{fig:outflow} yields a distance estimate of 145\,pc to IRC\,+10216.
This is in accord with modern estimates lying in the range 110 -- 170\,pc
(\cite{WDS94}; \cite{LeB97}; \cite{GVM98}).
However, there are too many uncertainties involved to place high confidence
in this result.
In addition to the geometric ambiguities already mentioned, it is unclear
if the dust clump followed here (NE2) has finished accelerating and is at its
terminal velocity.
Furthermore, the value of V$_\infty$ measured in the spherical molecular shell 
may not be a good measure of the inner dust motions.
It is difficult to imagine that any dramatic change from spherical outflow to 
an equatorial disk and bipolar lobes in IRC\,+10216 did not also entail changes 
in the velocity structure.

Taking a distance of 145\,pc, proper motions of 11.5, 17.8 \& 25.5\,mas.yr$^{-1}$ 
(Figure~\ref{fig:outflow}) imply outflow velocities of  
7.9, 12.2 \& 17.5\,km.sec$^{-1}$ for points NE1, EC1, and NE2 respectively.
As computed above, the velocity structure is consistent with an apparent 
uniform acceleration of $3.4\pm0.5$\,mas.yr$^{-2}$ from rest at $80\pm20$\,mas.
However, the projection of a three-dimensional motion onto the plane of the sky, 
and unknowns associated with the initial conditions of each clump may also
account for the range of observed flow speeds.
Models of radiatively-driven dust acceleration predict a more complicated 
velocity law dependent upon many properties of star and the outflowing gas 
and dust (\cite{Kwok75}; \cite{PP86}).
However, detailed comparison with such predictions is premature until 
the velocity of a single feature can be shown to be changing over time.

%___________________________________CONCLUSIONS________________________
\section{Conclusions}

Diffraction-limited images recovered using interferometric techniques from a
multi-epoch study spanning more than two years at the Keck~I telescope are
presented.
Taken in the near-infrared K~Band, the maps have revealed an asymmetric and
clumpy structure at angular resolutions exceeding the expected diameter of
the stellar photosphere ($\sim$ 50\,mas).
The most likely morphology for the circumstellar environment is an optically
thick circumstellar disk or torus, possibly tilted towards the line of 
sight in the South revealing the hot inner cavity and emission from the
stellar photosphere.
The angular separations of clumps of material thought to be in the Northern 
bipolar lobe have been followed over time, revealing increasing separation
from the compact core to the South.
Outflow velocities derived from this motion are consistent with estimates 
of the radial outflow velocity (from CO measurements) and the expected 
distance.
Clumps at greater distances from the Core were found to show increasing 
velocities, which may be taken as evidence for acceleration in the inner regions; 
the effects of geometrical projection; or the result of a past event which 
ejected material with a range of velocities.
In addition to the changing separations of components, the appearance of the
inner nebula was found to be evolving in other ways, however none were
interpreted as evidence for new dust condensation over the pulsation cycle.
Further modelling of this system is currently underway, and will be 
presented in a subsequent paper.

%___________________________________ACKNOWLEDGMENTS___________________
\acknowledgments

{We would like to thank the referee, Matt Bobrowsky, for helpful suggestions
and  Devinder Sivia for the maximum-entropy mapping program ``VLBMEM''. 
Data presented herein were
obtained at the W.M. Keck Observatory, which is operated as a
scientific partnership among the California Institute of Technology,
the University of California and the National Aeronautics and Space
Administration.  The Observatory was made possible by the generous
financial support of the W.M. Keck Foundation.  
This work is a part of a
long-standing interferometry program at U.C. Berkeley, supported by
the National Science Foundation (Grant AST-9315485 and AST-9731625) 
by the Office of Naval Research (OCNR N00014-89-J-1583), and the 
France-Berkeley Fund.}

%\pagebreak UNCOMMENT
%___________________________________BIBLIOGRAPHY_______________________
%\clearpage

\end{document}